\documentclass[sigconf]{acmart}

\usepackage{graphicx} 
\usepackage[utf8]{inputenc} 
\usepackage[T1]{fontenc}    
\usepackage[english]{babel}
\usepackage{booktabs}       
\usepackage{amsfonts}       
\usepackage{nicefrac}       
\usepackage{microtype}      
\usepackage{xcolor}         
\usepackage{enumitem}
\usepackage{colortbl}
\usepackage{multirow}
\usepackage{subcaption}
\usepackage{dsfont}
\usepackage{ulem}

\newcommand\light[1]{{\footnotesize \textit{\textcolor{gray}{#1}}}}
\newcommand\std[1]{{\small \textit{\textcolor{gray}{$\pm$#1}}}}
\newcommand\ul[1]{\underline{#1}}

\newcommand\bld[1]{\textbf{#1}}

\newcommand\SRpp{SASRec\texttt{++}}

\title{Closing the Performance Gap in Generative Recommenders with Collaborative Tokenization and Efficient Modeling}

\author{Simon Lepage}
\affiliation{%
    \institution{CRITEO AI Lab \& LIGM, École Nationale des Ponts et Chaussées}
    \city{Paris}
    \country{France}
}
\email{s.lepage@criteo.com}

\author{Jeremie Mary}
\affiliation{%
    \institution{CRITEO AI Lab}
    \city{Paris}
    \country{France}
}
\email{j.mary@criteo.com}

\author{David Picard}
\affiliation{
    \institution{LIGM, École Nationale des Ponts et Chaussées}
    \city{Marne-la-Vallée}
    \country{France}
}
\email{david.picard@enpc.fr}



\begin{document}
    
    \begin{abstract}
        Recent work has explored generative recommender systems as an alternative to traditional ID-based models, reframing item recommendation as a sequence generation task over discrete item tokens. While promising, such methods often underperform in practice compared to well-tuned ID-based baselines like SASRec. In this paper, we identify two key limitations holding back generative approaches: the lack of collaborative signal in item tokenization, and inefficiencies in the commonly used encoder-decoder architecture.
        To address these issues, we introduce COSETTE, a contrastive tokenization method that integrates collaborative information directly into the learned item representations, jointly optimizing for both content reconstruction and recommendation relevance. Additionally, we propose MARIUS, a lightweight, audio-inspired generative model that decouples timeline modeling from item decoding. MARIUS reduces inference cost while improving recommendation accuracy.
        Experiments on standard sequential recommendation benchmarks show that our approach narrows, or even eliminates, the performance gap between generative and modern ID-based models, while retaining the benefits of the generative paradigm.
    \end{abstract}
    
    \maketitle
    
    \section{Introduction}

Generative recommender systems have recently emerged as a promising alternative to traditional ID-based methods for sequential recommendation. Rather than ranking items based on similarity scores, generative approaches reframe item retrieval as a sequence generation task~\citep{rajput2023recommender}. In this paradigm, each item is tokenized into a semantic identifier, typically a tuple of discrete tokens, and the model autoregressively generates the identifier of the next item in the sequence. This formulation offers several advantages: it enables token-level alignment between user history and candidate items, leads to better generalization thanks to the Semantic IDs~\cite{singh24better}, removes the reliance on non-differentiable approximate nearest neighbor (ANN) modules, and eliminates large embedding tables that increasingly hinder the scalability of traditional systems~\citep{liu2024embedding, zhai2024Action}.

However, practical experience in industry suggests that well-tuned ID-based methods like SASRec~\cite{kang2018self} with simple modifications~\cite{TIGERReviews} can significantly outperform the results typically reported in the literature.
We identify two key limitations that contribute to the performance gap: (1) the lack of collaborative signal during the item tokenization process, and (2) inefficiencies in the architecture of current generative models.

The first issue arises because generative recommenders typically rely on residual vector quantization~\citep{zeghidour2021soundstream} trained to reconstruct frozen item embeddings derived from content (e.g., textual metadata)~\citep{rajput2023recommender, zhai2025multimodal, zheng2024adapting}. However, this objective is misaligned with the downstream goal of recommendation, which requires distinguishing between items based on collaborative signals rather than reconstructing their content. While some work explores contrastive objectives~\citep{zhu2024cost}, they often remain anchored to fixed, non-collaborative embeddings.
Others incorporate collaborative information by leveraging pretrained SASRec embeddings~\cite{wang2024learnable}, but this introduces an additional pre-training phase. 
To address this, we propose COSETTE (\textbf{CO}llaborative and \textbf{SE}mantic \textbf{T}okenization of \textbf{T}ext \textbf{E}mbeddings), a novel tokenization approach that directly integrates collaborative signals into the quantization process via a latent contrastive loss. 

The second limitation lies in the architecture of current generative models. Many recent systems adopt an encoder-decoder structure inspired by TIGER~\citep{rajput2023recommender}, where the semantic identifiers are concatenated before being fed to a bidirectional encoder, and decoded autoregressively~\citep{wang2024learnable, zhai2025multimodal, zhu2024cost}. This design suffers from significant inefficiencies such as long input sequences and the absence of causal structure in the encoder preventing KV-Caching. It also suffers from high computational cost in the decoder due to cross-attention over the entire input, a limitation that is particularly costly during top-$K$ generation, where the decoder must be queried repeatedly to generate new items.
We address these bottlenecks with MARIUS (\textbf{M}ulti-scale \textbf{A}ttention as \textbf{R}ecommendation \textbf{I}ndex with f\textbf{US}ion), an architecture inspired by the RQ-Transformer~\citep{lee2022autoregressive} used in audio modeling~\citep{yang2023uniaudio, kyutai2024moshi, zhu2024generative}. MARIUS separates temporal and local modeling into dedicated components, enabling efficient item generation while improving recommendation accuracy. This decoupled design facilitates caching, reduces inference latency, and improves scalability.

The goal of this paper is to close the gap between generative and ID-based recommender systems. We are able to achieve such a goal with the following contributions:
\begin{itemize}[leftmargin=1.8em]
    \item We introduce and evaluate a modernized SASRec variant on standard benchmarks, to establish a very strong baseline.
    \item We present COSETTE, a contrastive tokenization method that embeds collaborative signals directly into discrete semantic representations.
    \item We take inspiration from the audio modeling community and propose MARIUS, which decouples temporal modeling and item decoding for improved efficiency and accuracy.
\end{itemize}
    \section{Related Work}

    \paragraph{Sequential Recommendation.}
    Recommender systems are fundamental components of online platforms, designed to provide personalized content to individual users. Sequential recommendation, in particular, aims to model a user's evolving preferences by leveraging their historical interaction data to predict the next item of interest~\citep{He2023Survey, deldjoo2024review}. Early approaches in this domain relied on Markov chains and matrix factorization techniques to capture user behavior~\citep{He2016Fast}. With the advent of deep learning, more sophisticated models have emerged, utilizing architectures such as Convolutional Neural Networks~\citep{Tang2018Personalized}, Graph Neural Networks~\citep{wu2019session}, and Recurrent Neural Networks~\citep{hidasi2015session}. Among these, Transformer-based models~\citep{vaswani2017attention} have demonstrated notable performance gains, employing either causal~\citep{kang2018self} or masked~\citep{Sun2019BERT4REC} sequence modeling strategies. The majority of these methods adopt a retrieve-then-rank paradigm, wherein each item is represented by a learned embedding, enabling efficient Approximate Nearest Neighbor (ANN) search during inference (e.g., using FAISS~\citep{johnson2019billion} or a vector Database).

    The deployment of sequential recommendation systems on real-world data necessitates their scalability to large item catalogs, motivating their study through the lens of scaling laws~\citep{kaplan2020scaling}. Recent studies, such as \cite{shen2024optimizing} and \cite{zhang2024scaling}, report promising results for attention-based architectures akin to SASRec. However, scaling these models introduces significant computational challenges, as the embedding table grows linearly with the number of items, adversely affecting both training and inference efficiency. Addressing this limitation through techniques such as sharding strategies, architectural innovations, and specialized optimizers remains an active and critical area of research~\citep{liu2024embedding, shen2024optimizing, zhai2024Action}.

    \paragraph{Generative Recommender Systems.}
    Generative retrieval~\cite{Tay2022Transformer} has emerged as a promising paradigm for information retrieval, wherein models are trained end-to-end to directly generate item identifiers. This approach eliminates the need for large item embedding tables and non-differentiable ANN systems, enabling the modeling of more complex similarity functions beyond the limitations of dot-product-based retrieval. In the context of recommendation, two main approaches have been explored. The first leverages pre-trained Large Language Models (LLMs), capitalizing on their general knowledge, multitasking capabilities, and ability to engage in natural language interactions~\citep{Zhao2024Recommender, Wu2024Survey}. A central challenge in this setting is extending the LLM’s vocabulary to include item identifiers without compromising its core capabilities, typically addressed through multi-task prompt-based fine-tuning~\citep{Geng2022Recommendation, zheng2024adapting}. The second approach involves training dedicated sequence-to-sequence models. For instance, TIGER~\citep{rajput2023recommender} demonstrates strong performance by training a T5 model directly on item sequences. Subsequent works building on TIGER have continued to employ encoder-decoder architectures~\cite{zhu2024cost, wang2024learnable, zhai2025multimodal}. In this work, we explore the use of the RQ-Transformer~\citep{lee2022autoregressive}, inspired by its success in audio modeling~\citep{yang2023uniaudio, zhu2024generative, kyutai2024moshi}. We show that it is well suited for sequential recommendation, achieving superior performance with significantly lower computational cost, particularly during inference.    
        
    \paragraph{Semantic Tokenization}
    The generative recommendation paradigm typically follows a two-stage procedure, beginning with item tokenization to reduce vocabulary size and enable semantic sharing among related items~\cite{singh24better}. While some approaches use raw text tokens as general-purpose representations~\citep{Li2023Text}, the Semantic ID strategy encodes items as tuples of task-specific codes, leveraging the exponential capacity of hierarchical representations to achieve effective vocabulary compression~\cite{hua2023how}. A key technique in this context is RQ-VAE~\cite{zeghidour2021soundstream, lee2022autoregressive}, which applies multi-step residual quantization with a reconstruction objective, and serves as a foundational component in modern generative recommender systems~\citep{rajput2023recommender, wang2024learnable, zhu2024cost, zheng2024adapting}.
    Subsequent work has sought to enhance this mechanism: \cite{wang2024learnable} introduced regularization of codebook diversity via a constrained K-Means algorithm; \cite{zheng2024adapting} removed the final non-semantic deduplication token and applied post-training token balancing using the Sinkhorn-Knopp algorithm~\cite{Cuturi2013Sinkhorn}; \cite{zhai2025multimodal} extended the framework to multimodal semantics by incorporating image data; \cite{zhu2024cost} replaced the reconstruction objective with a contrastive loss to improve the discriminative quality of the latent space; and \cite{wang2024learnable, wang2024eager} incorporated collaborative signals into the quantization process by aligning token representations with embeddings from a pretrained model such as SASRec~\cite{kang2018self}.
    In this work, we introduce COSETTE, a novel approach that uses contrastive learning to integrate collaborative signals directly into the tokenization process, achieving improved performance without pretrained models.

    \section{Method}

    We address two key limitations of generative recommenders: misaligned item tokenization that does not capture collaborative signals, and inefficient sequence modeling architectures. To address the first issue, we propose \textbf{COSETTE} in Section~\ref{method:cosette}, a contrastive tokenization method that integrates both semantic and collaborative signals. To address the second issue, we propose \textbf{MARIUS} in Section~\ref{method:marius}, a lightweight generative model that decouples sequence modeling from item decoding to improve efficiency and accuracy.

    \subsection{Collaborative and Semantic Tokenization}\label{method:cosette}
\begin{figure}[t]
                \centering
                \includegraphics[width=\linewidth]{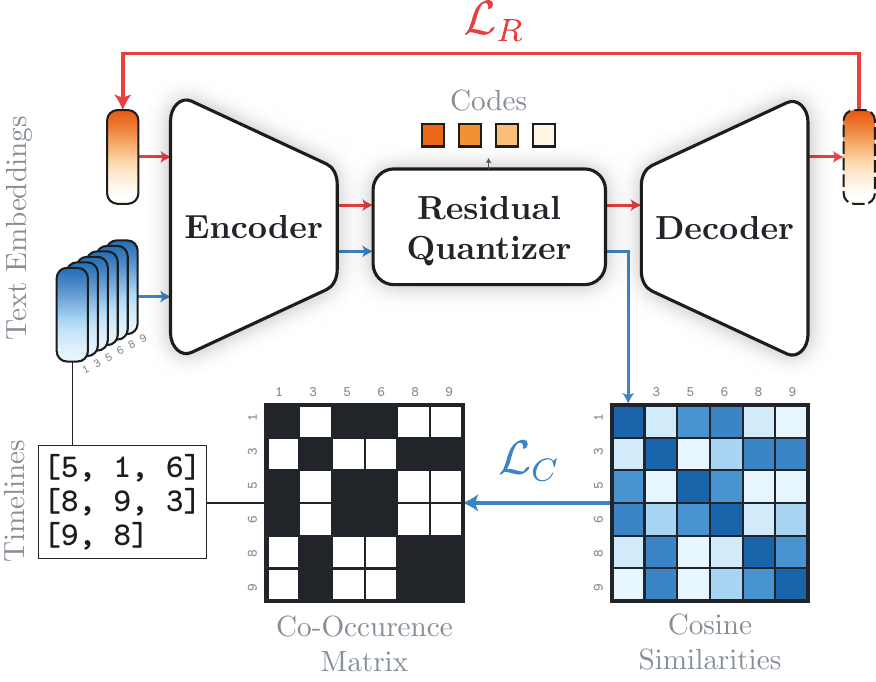}
                \caption{Overview of COSETTE. $\mathcal{L}_R$ encourages accurate reconstruction of input text embeddings after compression via a residual quantizer. $\mathcal{L}_C$ aligns similarities in the quantizer's latent space with the co-occurrence structure derived from batches of user timelines.}
                \label{fig:method_cosette}
                \Description[Diagram of COSETTE]{The diagram illustrates the architecture of the COSETTE tokenization model. On the left, input text embeddings and user timelines are processed by an encoder. The encoder outputs are passed to a residual quantizer, which produces discrete code representations and quantized latent representations. These representations are decoded to reconstruct the original embeddings, with a reconstruction loss L_R. Simultaneously, timelines are converted into a co-occurrence matrix, which is compared with a cosine similarity matrix of the quantized embeddings. This comparison yields a contrastive loss L_C, used to align collaborative signals with semantic similarities. The overall pipeline integrates both semantic and collaborative information into the learned codes.}
            \end{figure}

        COSETTE builds on the RQ-VAE framework~\citep{lee2022autoregressive} by incorporating collaborative signals through a contrastive learning objective. This enhances the learned item codes with information relevant to downstream recommendation tasks. Figure~\ref{fig:method_cosette} presents an overview of the method.

        \subsubsection{Reconstructive Quantization}
            Given an item and its metadata such as title, brand, price, or description, we first extract a semantic embedding $\boldsymbol{e}$ using a pre-trained language model (e.g., Sentence-T5-XL~\citep{ni2022sentence}). This embedding is then compressed into a latent vector $\boldsymbol{z} = \text{Encoder}(\boldsymbol{e})$.
            To discretize this latent representation, we apply residual quantization over $L$ levels, each associated with a codebook. Specifically, for each level $l \in \{1, \ldots, L\}$, we define a codebook $\mathcal{C}^l = \{\boldsymbol{c}_k^l\}_{k=1}^K$, where each $\boldsymbol{c}_k^l$ is a learnable code vector, and $K$ is the codebook size. The codebooks are initialized using K-Means clustering on the encoded latent vectors from the first training batch.
            Quantization proceeds iteratively as:
            \begin{align}
                    v_i &= \underset{k}{\text{argmin}} \lVert \boldsymbol{r}_i - \boldsymbol{c}_k^i \rVert_2^2 , \\
                    \boldsymbol{r}_{i+1} &= \boldsymbol{r}_i - \boldsymbol{c}_{v_i}^i ,
            \end{align}
            with the initial residual set to $\boldsymbol{r}_1 = \boldsymbol{z}$.         
            The quantized latent embedding is then computed as $\hat{\boldsymbol{z}} = \sum_{l=1}^L \boldsymbol{c}_{v_l}^l$, which is decoded to reconstruct the original semantic embedding as $\hat{\boldsymbol{e}} = \text{Decoder}(\hat{\boldsymbol{z}})$.
            The sequence of indices $\{v_i\}_{i=1}^L$ obtained through residual quantization forms a discrete code tuple that serves as the item’s identifier during downstream modeling. The coarse-to-fine structure of residual quantization naturally induces a tree-structured item space, which is well-suited for autoregressive generation.
    
            The training objective consists of two components: a quantization loss that aligns the residuals and code vectors, and a reconstruction loss that preserves semantic fidelity:
            \begin{align}\label{eq:LQ}
                \mathcal{L}_Q &= \sum_{l=1}^L \lVert\text{sg}[\boldsymbol{r}_l] - \boldsymbol{c}_{v_l}^l \rVert_2^2 + \beta \lVert \boldsymbol{r}_l - \text{sg}[\boldsymbol{c}_{v_l}^l]\rVert_2^2 ,\\
                \mathcal{L}_R &= \lVert \boldsymbol{e} - \hat{\boldsymbol{e}} \rVert_2^2 ,
            \end{align}
            where $\text{sg}[\cdot]$ denotes the stop-gradient operator, and we adopt the straight-through estimator~\cite{van2017neural} for backpropagation through the discrete code selections.

        \subsubsection{Collaborative Quantization}
            The semantic quantization described above does not leverage collaborative signals present in user interaction sequences, as each item is processed independently. Prior work has demonstrated that collaborative indexing methods such as those based on spectral clustering over item co-occurrence matrices can significantly improve recommendation performance~\citep{hua2023how}. While some recent approaches inject collaborative information by pre-training item embeddings on interaction data~\citep{wang2024learnable, wang2024eager}, we instead integrate this signal directly into the tokenization via contrastive learning. Specifically, we use item co-occurrence statistics to guide the learning of discrete codes through a contrastive objective.
    
            We start by sampling a batch of user timelines $\mathcal{T} = \{T_i\}_{i=1}^B$ where each $T_i$ a sequence of item interactions. From these, we extract the set of unique items appearing across all sequences $\{t_i\}_{i=1}^{B'}$. For each item, we retrieve its semantic embedding $\boldsymbol{e}_i$, and use the encoder and residual quantizer to compute the corresponding quantized latent representation $\hat{\boldsymbol{z}}_i$.
    
            We then apply a contrastive loss over item pairs to encourage items that co-occur in timelines to have similar representations. Following the Sigmoid Contrastive Loss formulation~\citep{zhai2023sigmoid}, we treat each item pair as a binary classification task:
            \begin{align}
                \mathcal{L}_C &= \frac{1}{B'}
                    \sum_{i=1}^{B'}\frac{1}{Y_i}
                    \sum_{j=1}^{B'}
                    \text{log}\left({1 + e^{y_{ij}(-t\langle\hat{\boldsymbol{z}}_i,\hat{\boldsymbol{z}}_j\rangle + b)} }\right), \\
                \text{where} \quad y_{ij} &= \begin{cases}
                    1 \quad \text{if} \quad \exists T\in \mathcal{T} : t_i \in T \land t_j \in T ,\\
                    -1 \quad \text{otherwise.}
                \end{cases}
            \end{align}
            Here, $\langle \cdot, \cdot \rangle$ denotes cosine similarity, and $Y_i = \sum_{j=1}^{B'} \mathds{1}_{y_{ij} = 1}$ normalizes the number of positives for item $i$. We parameterize the temperature as $t = e^{t'}$, with $t'$ and the bias $b$ are learnable.
            To preserve quantization quality during contrastive training, we additionally compute a quantization loss $\mathcal{L}_{Q_C}$ on the batch of collaboratively supervised items. This loss is defined identically to $\mathcal{L}_Q$ (Equation \ref{eq:LQ}), but applied to the residuals and codebooks computed in this contrastive batch. Our final loss is
            \begin{equation}
                \mathcal{L}_{\text{COSETTE}} = \mathcal{L}_Q + \mathcal{L}_R + \mathcal{L}_{Q_C} + \lambda \mathcal{L}_C,
            \end{equation}
            with $\lambda$ a normalization weight to put $\mathcal{L}_C$ in the same range as the other losses.
            \subsubsection{Post-processing.}
            Despite the exponential number of possible combinations in residual quantization ($K^L$), item collisions can still occur. In practice, we observe that the contrastive training in COSETTE reduces the rate of such collisions (see Figure~\ref{fig:collisions}), but does not eliminate them entirely. This poses a challenge for retrieval, as items sharing the same code become indistinguishable.
    
            Two main strategies have been proposed to address this issue. The first, used in prior work~\citep{rajput2023recommender, hua2023how}, appends an additional disambiguating token that does not carry semantic meaning. The second, introduced in LC-Rec~\citep{zheng2024adapting}, imposes a uniformity constraint during training to discourage collisions, and is further extended in MQL4GRec~\citep{zhai2025multimodal}, which applies a post-training reallocation procedure. This reassigns colliding items to nearby unused code tuples based on residual distances to codebook vectors.
    
            We adopt this second approach in COSETTE, as we find it more effective for maintaining semantic and collaborative integrity while minimizing collisions.

    \subsection{Generative Model}\label{method:marius}
        \begin{figure}[t]
            \centering
            \includegraphics[width=\linewidth]{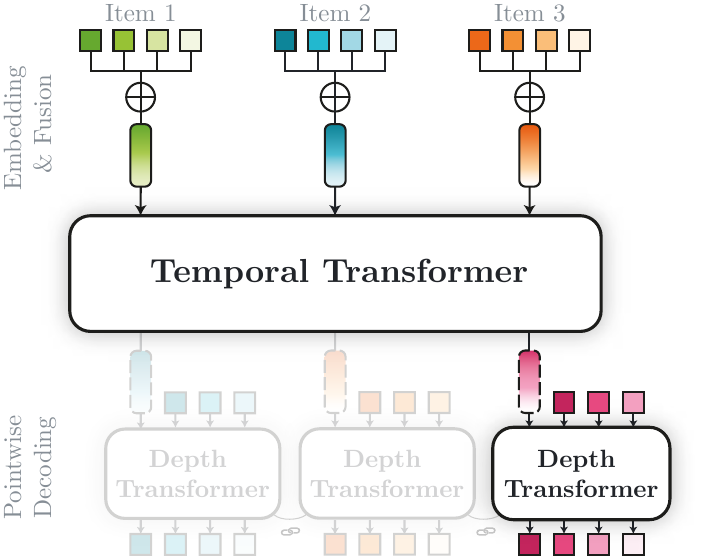}
            \caption{Overview of MARIUS. For each item in a sequence, we fuse the embeddings of the associated codes, and apply an autoregressive Temporal Transformer. The Depth Transformer is applied pointwise and autoregressively predicts the codes of the next item.}
            \label{fig:method_marius}
            \Description[Diagram of MARIUS]{The diagram illustrates the MARIUS generative model. Each item in a user sequence is represented by a group of discrete codes, which are embedded and fused into a single vector. These fused item representations are input sequentially into a Temporal Transformer, which captures the sequence-level context. The output of the Temporal Transformer is then passed to a Depth Transformer, which autoregressively decodes the individual codes of the next item in the sequence. The process is repeated pointwise for each prediction step.}
        \end{figure}
        
        Most prior generative recommender systems rely on encoder-decoder architectures inspired by the T5 model, operating over sequences of item semantic IDs. While effective, these architectures are computationally expensive (see cost analysis in Section~\ref{sec:efficiency}). In contrast, the RQ-VAE framework introduced the RQ-Transformer~\citep{lee2022autoregressive}, which has gained traction in recent audio modeling research~\citep{yang2023uniaudio, zhu2024generative, kyutai2024moshi}, is much more lightweight. Our proposed model, MARIUS, builds on this foundation. Figure~\ref{fig:method_marius} shows an overview of the architecture.

        Given a timeline $T = [t_1, \ldots, t_N]$, we construct a matrix $V\in \mathbb{N}^{N\times L}$ where each row contains the discrete code indices for an item. We define a learnable embedding matrix $E\in\mathbb{R}^{K\cdot L \times d}$, and represent each item as a sum over its $L$ code-level embeddings as $\boldsymbol{x}_i = \sum_{j=1}^L E_{V_{ij}}$.
        We apply a \textit{Temporal Transformer} over the item sequence $[\boldsymbol{x}_i]_{i=1}^N$ to compute a temporal context vector for each step $1 < s \le N$
        \begin{equation}
            \boldsymbol{h}_s = T_T(\boldsymbol{x}_1, \ldots, \boldsymbol{x}_{s-1}) \in \mathbb{R}^d ,
        \end{equation}
        which is then projected to match the input dimensionality of the \textit{Depth Transformer} as $\boldsymbol{h}'_s=Lin(\boldsymbol{h}_s) \in \mathbb{R}^{d'}$. 
        The Depth Transformer operates autoregressively over the code dimensions of the next item. For each code level $1 \le k \le L$, it predicts the logits estimates 
        \begin{equation}
            \boldsymbol{l}_{s,k} = T_D(\boldsymbol{h}'_s, E'_{V_{s,1}}, \ldots, E'_{V_{s,k-1}}) \in \mathbb{R}^{L\cdot K} .
        \end{equation}
        
        This two-stage design decouples sequence modeling from item decoding: the Temporal Transformer always processes $N$ tokens, one per item, rather than $NL$ in typical generative recommender systems.The Depth Transformer handles only $L$ tokens per prediction, without cross-attention to the entire sequence.
        Intuitively, the Temporal Transformer plays a role similar to that of SASRec, summarizing the sequence-level context, while the Depth Transformer replaces the role of an item index, learning to generate discrete item codes autoregressively.
    \section{Experiments}
    \subsection{Experimental Settings}
        \paragraph{Datasets.} We conduct our experiments on the Amazon Reviews datasets \citep{mcauley2015image, hou2023vqrec}, which are large collections of ratings grouped by verticals. The 2014 version is a common benchmark for Sequential Recommendation research but is limited in size, with only one vertical containing more than 70k products after preprocessing. To benchmark our models on larger and more realistic datasets, we use the 2023 version~\cite{hou2024bridging}, which are an order of magnitude larger. Following prior work, we filter out items and users that appear fewer than five times. Dataset statistics are provided in Table~\ref{tab:datasets}.
        
        \paragraph{Evaluation.} We report the widely used R@k and NDCG@k metrics in the main tables and focus on R@10 for smaller experiments and ablations. We adopt the leave-last-out strategy to define validation and test splits, using the penultimate item for validation and the last item for testing. For ablations where direct comparison with prior work is not required, we report only validation metrics to avoid biasing the main comparisons.

        \begin{table}[t]
            \centering
            \caption{Datasets statistics after 5-core filtering.}
            \resizebox{\linewidth}{!}{%
                \begin{tabular}{cr|ccc}
   \toprule
    & Name & \#Users & \#Items & \#Interactions \\
   \midrule
   \parbox[t]{2mm}{\multirow{5}{*}{\rotatebox[origin=c]{90}{\bld{2014}}}}
   & Beauty             &  22,363  & 12,101 & 198,502   \\
   & Movies \& TV       &  123,960 & 50,052 & 1,697,533 \\
   & Sports \& Outdoors &  35,598  & 18,357 & 296,337   \\
   & Toys \& Games      &  19,412  & 11,924 & 167,597   \\
   & Video Games        &  24,303  & 10,762 & 231,780   \\
   \midrule
   \parbox[t]{2mm}{\multirow{10}{*}{\rotatebox[origin=c]{90}{\bld{2023}}}}
   & Arts Crafts \& Sewing   &  197,286 & 89,958  & 1,786,437 \\
   & Automotive              &  632,138 & 267,275 & 6,072,233 \\
   & Beauty \& Personal Care &  729,576 & 207,649 & 6,624,441 \\
   & Cell Phones \& Accessories&  380,999 & 111,480 & 2,752,785 \\
   & Grocery \& Gourmet Food &  404,760 & 132,865 & 3,948,741 \\
   & Health \& Household     &  796,054 & 184,346 & 7,176,552 \\
   & Office Products         &  223,308 & 77,551  & 1,800,878 \\
   & Patio Lawn \& Garden    &  416,649 & 133,447 & 3,441,178 \\
   & Pet Supplies            &  594,800 & 114,625 & 5,277,315 \\
   & Sports \& Outdoors      &  409,772 & 156,235 & 3,472,020 \\
   \bottomrule
\end{tabular}
            }
            \label{tab:datasets}
        \end{table}
    
    \subsection{\SRpp{}}

        SASRec~\cite{kang2018self} is a transformer-based sequential recommendation model, typically trained with a binary cross-entropy loss to predict the next item. In practice, the official implementation uses a single negative item, and applies a filtering step to exclude items already present in the query timeline from the returned logits. Subsequent works, such as S3-Rec~\citep{zhou2020S3Rec}, adopt the same loss and are frequently used as baselines in the Sequential Recommendation community. However, recent studies on SASRec's scaling~\citep{shen2024optimizing, zhai2023revisiting, zhang2024scaling} favor Softmax and Sampled-Softmax losses over the full vocabulary for improved stability and performance~\cite{klenitskiy2023turning}. 
        
        We adopt similar modifications, along with random timeline cropping and shuffling of items sharing the same timestamp, and denote the result \SRpp{} in the rest of the paper. Results in Table~\ref{tab:sasrec_variants} show a $+50\%$ improvement in R@10 compared to the commonly reported method. These augmentations show that SASRec is mostly limited by the training setup. Using a more appropriate training allows SASRec to reach very competitive performances on small scale datasets and looks very promising at large scales. For large datasets, we find that using a sampled InfoNCE loss~\citep{sohn_improved_2016, oord2018representation} with a temperature of $0.05$ yields better and more stable results.
        
        \begin{table}[t]
            \centering
            \caption{Mean and standard deviation of test R@10 for SASRec variants, over 5 runs. \textit{SASRec+Filtering} is equivalent to the official SASRec implementation. Our final implementation brings a relative uplift of over +50\%.}
            \begin{tabular}{r|ccc}
   \toprule
    & Beauty & Sports & Toys \\ 
   \midrule
   SASRec BCE        & 5.48\% \std{0.05} & 3.81\% \std{0.08} & 5.86\% \std{0.08} \\
   + Filtering       & 6.33\% \std{0.14} & 4.19\% \std{0.09} & 6.49\% \std{0.14} \\
   + Cross-Entropy   & 8.70\% \std{0.11} & 5.53\% \std{0.04} & 8.84\% \std{0.10} \\
   + Crop \& Shuffle & \bld{9.73\%} \std{0.10} & \bld{6.44\%} \std{0.10} & \bld{9.94\%} \std{0.06} \\
   \bottomrule
\end{tabular}
            \label{tab:sasrec_variants}
        \end{table}

    \subsection{COSETTE}
        We begin by analyzing COSETTE’s design through a series of ablations on its loss components and key training parameters, focusing on stability and generalization. We then compare it to five existing quantization methods on datasets containing 10k, 50k, and 114k items, highlighting its consistent performance gains across different datasets scales and generative recommender architectures. The model sizes for each dataset are presented in Table~\ref{tab:ablation_model_sizes}.

        \subsubsection{Ablations}

            To understand the contribution of each component in COSETTE's training objective, we conduct an ablation study by modifying one loss term at a time, and report the results in Table~\ref{tab:cosette_loss}.
            The full loss combines a quantization loss, a reconstruction loss, and a collaborative loss. We observe that replacing the reconstruction loss by the collaborative loss $\mathcal{L}_C$ already yields improvements on the two largest datasets, and similar results for the smallest one. Applying the reconstruction loss on the same items as those sampled for the collaborative loss (i.e. popularity-weighted sampling), denoted $\mathcal{L}_{R_P}$, degrades the results for the small dataset. Applying the reconstruction loss on a set of uniformly sampled items, denoted  $\mathcal{L}_{R_U}$, improves the results for all datasets, in particular the smallest one.
            These results confirm that both collaborative alignment and faithful reconstruction are necessary for high-quality item representations.

            Beyond loss components, we analyze a few hyperparameters that helped stabilize training and reduce overfitting. We detail the results in Appendix~\ref{app:cosette_params}, and highlight key findings here: 
            \begin{itemize}[leftmargin=1.8em]
                \item \bld{Loss weight.} The reconstruction and quantization losses are MSE-based, whereas the contrastive loss is a cross-entropy term, leading to different scales. We find that setting $\lambda = 10^{-3}$ appropriately balances these components.
                \item \bld{Temperature and Bias.} We set the temperature and the bias of the loss as learnable parameters. We initialize them with $t'=2$ and $b=-8$, close to the values reached toward the end of the trainings.
                \item \bld{Batch size and Steps.} For larger datasets, we use a batch size of 1024 and observe that downstream performance continues to improve up to around 500k steps. On smaller datasets, reducing both batch size and training steps acts as a form of regularization to avoid overfitting.
                \item \bld{Dropout.} We apply a small dropout rate of $0.1$ on the projection layers when training on smaller datasets, which helps generalization without degrading code quality.
            \end{itemize}
            Additionally, we find that the contrastive nature of COSETTE further reduces the number of collisions (see Figure~\ref{fig:collisions} in the Appendix).
            
            \begin{table}[t]
                \centering
                \caption{Chosen model sizes for COSETTE ablations, based on the insights from Appendix~\ref{app:model_sizing}.}
                \resizebox{\linewidth}{!}{%
                    \begin{tabular}{r|cc}
    \toprule
     & TIGER & MARIUS \\
     \midrule
     Video Games  & $d=128$, $L_E=4$, $L_D=4$ & $d=256$, $L_T=2$, $L_D=2$ \\
     Movies \& TV & $d=128$, $L_E=2$, $L_D=6$ & $d=512$, $L_T=4$, $L_D=4$ \\
     Pet Supplies & $d=128$, $L_E=2$, $L_D=8$ & $d=512$, $L_T=4$, $L_D=6$ \\
     \bottomrule
\end{tabular}
                }
                \label{tab:ablation_model_sizes}
            \end{table}
        
            \begin{table}[t]
                \centering
                \caption{Ablation of components of the COSETTE loss. We report the mean R@10 and standard deviation over 5 runs. For \textit{Movies \& TV} and \textit{Pet Supplies}, the collaborative signal alone already improves the results. We obtain the best results when combining it with the uniform reconstruction loss.}
                \resizebox{\linewidth}{!}{%
                    \begin{tabular}{cccc|ccc}
    \toprule
            $\mathcal{L}_Q$ & $\mathcal{L}_{R_U}$ & $\mathcal{L}_C$ & $\mathcal{L}_{R_P}$ & Video Games & Movies \& TV & Pet Supplies \\
    \midrule
    \checkmark & \checkmark &            &            & 14.84\% \std{0.08} & 9.30\% \std{0.02} & 5.00\% \std{0.02} \\
    \checkmark &            & \checkmark &            & 14.83\% \std{0.03} & 9.85\% \std{0.03} & 5.21\% \std{0.01} \\
    \checkmark &            & \checkmark & \checkmark & 14.66\% \std{0.07} & 9.86\% \std{0.04} & 5.30\% \std{0.01} \\
    \checkmark & \checkmark & \checkmark &            & \bld{15.02\%} \std{0.06} & \bld{9.90\%} \std{0.05} & \bld{5.32\%} \std{0.01} \\
    \bottomrule
\end{tabular}

                }
                \label{tab:cosette_loss}
            \end{table}

        \subsubsection{Comparison with other quantization methods} \label{sec:cosette_comp}
        \begin{table*}[t]
            \centering
            \caption{Comparison of Generative and ID-based recommenders on Amazon Reviews 2014. While TIGER-based approaches actually underperform compared to a strong ID-based baseline, our proposed method bridges the gap. For the models that we trained (CoST, LETTER and MARIUS), we report the mean and standard deviation over 5 runs. \bld{Bold} numbers are the best, \underline{Underline} are within standard deviation. $^\dagger$ : taken from~\cite{rajput2023recommender, wang2024eager}.}
            \resizebox{\textwidth}{!}{%
                \begin{tabular}{l|cccc|cccc|cccc}
    \toprule
    & \multicolumn{4}{c|}{\textbf{Beauty}} & \multicolumn{4}{c|}{\textbf{Sports \& Outdoors}} & \multicolumn{4}{c}{\textbf{Toys \& Games}}  \\
    Method & R@5 & NDCG@5 & R@10 & NDCG@10 & R@5 & NDCG@5 & R@10 & NDCG@10 & R@5 & NDCG@5 & R@10 & NDCG@10 \\
    \midrule
    Caser$^\dagger$~\cite{Tang2018Personalized}    & 2.05\% & 1.31\% & 3.47\% & 1.76\% & 1.16\% & 0.72\% & 1.94\% & 0.97\% & 1.66\% & 1.07\% & 2.70\% & 1.41\% \\ 
    HGN$^\dagger$~\cite{ma2019Hierarchical}      & 3.25\% & 2.06\% & 5.12\% & 2.66\% & 1.89\% & 1.20\% & 3.13\% & 1.59\% & 3.12\% & 2.21\% & 4.97\% & 2.77\% \\ 
    GRU4Rec$^\dagger$~\cite{hidasi2015session}     & 1.64\% & 0.99\% & 2.83\% & 1.37\% & 1.29\% & 0.86\% & 2.04\% & 1.10\% & 0.97\% & 0.59\% & 1.76\% & 0.84\% \\ 
    BERT4Rec$^\dagger$~\cite{Sun2019BERT4REC}      & 2.03\% & 1.24\% & 3.47\% & 1.70\% & 1.15\% & 0.75\% & 1.91\% & 0.99\% & 1.16\% & 0.71\% & 2.03\% & 0.99\% \\ 
    FDSA$^\dagger$~\cite{Zhang2019Feature}         & 2.67\% & 1.63\% & 4.07\% & 2.08\% & 1.82\% & 1.22\% & 2.88\% & 1.56\% & 2.28\% & 1.40\% & 3.81\% & 1.89\% \\ 
    SASRec$^\dagger$~\cite{kang2018self}           & 3.87\% & 2.49\% & 6.05\% & 3.18\% & 2.33\% & 1.54\% & 3.50\% & 1.92\% & 4.63\% & 3.06\% & 6.75\% & 3.74\% \\ 
    TDM$^\dagger$~\cite{Zhu2018Learning}           & 4.42\% & 3.23\% & 6.38\% & 3.76\% & 1.27\% & 0.96\% & 2.21\% & 1.10\% & 3.05\% & 2.14\% & 3.59\% & 2.30\% \\ 
    S$^3$-Rec$^\dagger$~\cite{zhou2020S3Rec}       & 3.87\% & 2.44\% & 6.47\% & 3.27\% & 2.51\% & 1.61\% & 3.85\% & 2.04\% & 4.43\% & 2.94\% & 7.00\% & 3.76\% \\ 
    RecForest$^\dagger$~\cite{feng2022recommender} & 4.70\% & 3.41\% & 6.64\% & 4.00\% & 1.49\% & 1.01\% & 2.47\% & 1.33\% & 3.13\% & 2.60\% & 3.83\% & 2.85\% \\ 
    \midrule
    SASRec\texttt{++} & \bld{6.66\% \std{0.08}} & \bld{4.58\% \std{0.08}} & 9.73\% \std{0.10} & \bld{5.57\% \std{0.04}} &
                        \bld{4.37\% \std{0.09}} & \bld{2.96\% \std{0.05}} & 6.44\% \std{0.10} & \bld{3.62\% \std{0.04}} & 
                        \bld{7.03\% \std{0.05}} & \bld{4.93\% \std{0.03}} & \bld{9.94\% \std{0.06}} & \bld{5.86\% \std{0.02}}  \\
    \midrule
    TIGER$^\dagger$~\cite{rajput2023recommender} & 4.41\% \std{0.07} & 3.09\% \std{0.06} & 6.42\% \std{0.09} & 3.74\% \std{0.06} & 2.78\% \std{0.07} & 1.89\% \std{0.04} & 4.19\% \std{0.1} & 2.34\% \std{0.05} & 5.18\% \std{0.06} & 3.75\% \std{0.04} & 6.98\% \std{0.13} & 4.33\% \std{0.05}\\ 
    CoST-TIGER~\cite{zhu2024cost} & 4.46\% \std{0.05} & 2.98\% \std{0.03} & 6.88\% \std{0.13} & 3.75\% \std{0.05} &
                                    2.46\% \std{0.11}  & 1.58\% \std{0.06} & 3.98\% \std{0.14}  & 2.07\% \std{0.07}  &
                                    4.58\% \std{0.08}  & 3.11\% \std{0.07}  & 6.83\% \std{0.16}  & 3.83\% \std{0.10} \\
    LETTER-TIGER~\cite{wang2024learnable} & 4.73\% \std{0.13} & 3.11\% \std{0.12} & 7.39\% \std{0.14} & 3.97\% \std{0.12} &
                                            2.75\% \std{0.06} & 1.80\% \std{0.04} & 4.36\% \std{0.07} & 2.32\% \std{0.04} &
                                            4.54\% \std{0.11} & 3.06\% \std{0.09} & 6.84\% \std{0.03} & 3.80\% \std{0.07} \\
    EAGER$^\dagger$~\cite{wang2024eager} & 6.18\% & \ul{4.51\%} & 8.36\% & 5.25\% & 2.81\% & 1.84\% & 4.41\% & 2.36\% & 2.65\% & 1.77\% & 4.53\% & 5.05\% \\
    MARIUS (RQ-VAE) & 6.51\% \std{0.07} & 4.38\% \std{0.04} & 9.71\% \std{0.07} & 5.41\% \std{0.03} & 
                           3.97\% \std{0.04} & 2.59\% \std{0.03} & 6.10\% \std{0.07} & 3.27\% \std{0.03} & 
                           6.04\% \std{0.08} & 4.05\% \std{0.09} & 9.13\% \std{0.05} & 5.05\% \std{0.06} \\
    MARIUS (COSETTE) & \ul{6.58\% \std{0.09}} & 4.35\% \std{0.07} & \bld{10.02\% \std{0.08}} & 5.46\% \std{0.05} &
                            \ul{4.31\% \std{0.08}} & 2.83\% \std{0.06} & \bld{6.72\% \std{0.08}} & \bld{3.62\% \std{0.06}} & 
                            6.31\% \std{0.10} & 4.22\% \std{0.08} & 9.51\% \std{0.07} & 5.25\% \std{0.08}  \\
    \bottomrule
\end{tabular}
            }
            \label{tab:AR14}
        \end{table*}

            \begin{table}[t]
                \centering
                \caption{Comparison of text-based quantization methods. We train TIGER and MARIUS of the given size for each setting and report the validation R@10. $\dagger$: original author implementation. $^*$: our reimplementation.}
                \resizebox{\linewidth}{!}{
                    \begin{tabular}{cr|ccc}
    \toprule
            & & Video Games & Movies \& TV & Pet Supplies \\
    \midrule
    \parbox[t]{2mm}{\multirow{6}{*}{\rotatebox[origin=c]{90}{\bld{TIGER}}}}
    & PCA + RK                                      & 9.49\%  & 4.82\% & 2.69\% \\
    & RQ-VAE$^\dagger$~\cite{lee2022autoregressive} & 9.79\%  & 5.03\% & 2.57\% \\
    & \rotatebox[origin=c]{180}{$\Lsh$} no col.$^\dagger$ & 9.97\%  & 4.84\% & 2.71\% \\
    & CoST$^*$~\cite{zhu2024cost}                   & 9.58\%  & 4.85\% & 2.73\% \\
    & LETTER$^\dagger$~\cite{wang2024learnable}     & 10.33\% & 5.05\% & 2.48\% \\
    & COSETTE \textit{(Ours)}                       & \ul{10.89\%} & \ul{6.44\%} & \ul{3.30\%} \\
    \midrule
    \parbox[t]{2mm}{\multirow{6}{*}{\rotatebox[origin=c]{90}{\bld{MARIUS}}}}
    & PCA + RK                                         & 14.11\%     &  9.00\%      & 4.97\% \\
    & RQ-VAE$^\dagger$~\cite{lee2022autoregressive}    & 14.24\%     &  9.16\%      & 5.01\% \\
    & \rotatebox[origin=c]{180}{$\Lsh$} no col.$^\dagger$ & 14.84\%     &  9.30\%      & 5.00\% \\
    & CoST$^*$~\cite{zhu2024cost}                      & 14.74\%     &  9.24\%      & 5.06\% \\
    & LETTER$^\dagger$~\cite{wang2024learnable}        & 14.84\%     &  9.17\%      & 4.77\% \\
    & COSETTE \textit{(Ours)}                          & \bld{15.02} & \bld{9.90\%} & \textbf{5.32\%} \\
    \bottomrule
\end{tabular}
                }
                \label{tab:quantization}
            \end{table}

            We compare COSETTE with other quantization methods proposed in prior generative recommendation studies for generating semantic IDs:
            \begin{enumerate}[leftmargin=1.8em]
                \item \bld{PCA + Residual K-Means} applies PCA to project the data before using residual K-Means, similar to RQ-VAE.
                \item \bld{RQ-VAE} is the default quantization approach used in TIGER~\citep{rajput2023recommender}, trained with both quantization and reconstruction losses.
                \item \bld{RQ-VAE \textit{no col.}} follows MQL4GRec~\citep{zhai2025multimodal} and LC-Rec~\cite{zheng2024adapting}, training RQ-VAE with four quantization layers and using the Sinkhorn-Knopp algorithm~\citep{Cuturi2013Sinkhorn} to balance the final assignments. A post-training step remaps the last token to ensure uniqueness while minimizing the distance to the assigned centroid.
                \item \bld{CoST}~\citep{zhu2024cost} replaces the RQ-VAE reconstruction loss with a contrastive loss. We find that setting a lower $\alpha$ ($10^{-3}$) to better balance the losses improves performance. 
                \item \bld{LETTER}~\citep{wang2024learnable} addresses common shortcomings of RQ-VAE by balancing codebook assignments and incorporating a collaborative filtering signal. Balance is enforced via constrained K-Means regularization, while the collaborative signal is integrated through contrastive training against pretrained SASRec embeddings.
            \end{enumerate}

            Since all methods are based on the RQ-VAE framework, we align as many parameters as possible to isolate each method’s specific contributions. We initialize item embeddings using Sentence-T5-XL~\citep{ni2022sentence}, and project them to dimension 128 via an MLP with hidden dimensions 512 and 256, using ReLU activations. All models use codebooks of size 256 and represent items as tuples of length 4. These tuples are generated either through three quantization levels with an additional deduplication token (methods 1, 2, 4), or four levels with post-training reallocation to ensure uniqueness (methods 3, 5). We set equal weights for reconstruction and quantization losses, and use $\beta = 0.25$.

            Overall, we find that COSETTE yields significant improvements across all datasets and benefits both TIGER and MARIUS. COSETTE improves over its closest competitor LETTER, which highlights that using a raw collaborative signal seems to be more effective than aligning the codebook on pretrained collaborative features that may be incompatible with the quantized nature of the encoding.

    \subsection{MARIUS}
        We begin by exploring the design space of the MARIUS architecture, then evaluate its performance on small, standard benchmarks. We next highlight its strengths on large-scale benchmarks, an order of magnitude bigger than those commonly used, before concluding with a comparison of its computational complexity to TIGER.

        \subsubsection{Model Design and Sizing}
            We study the impact of key architectural choices in MARIUS (hidden dimension, number of layers, and dropout) by training multiple model variants on datasets of varying scale and evaluating their validation performance. Due to space constraints, detailed tables and figures are included in Appendix~\ref{app:model_sizing} and Table~\ref{tab:ar14_ablation}.

            We find that the \textit{Temporal Transformer} behaves similarly to SASRec: it processes the sequence using a single token per item, benefits from a high dropout rate ($0.4$), and does not require significant scaling with dataset size, as the complexity of user behavior and average sequence length remains stable.
            In contrast, MARIUS benefits from a larger hidden dimension, as it fuses the components of each semantic ID tuple into a single token.
            Additionally, the \textit{Depth Transformer}, responsible for decoding the output feature into a growing vocabulary of item tuples, requires increased capacity and must be scaled accordingly as the dataset size grows. This highlights the key dimensional difference of MARIUS compared to TIGER, with MARIUS processing shorter sequences of larger dimensions.

        \subsubsection{Performance on Standard Benchmarks}
            \begin{figure*}[t]
                \centering
                \includegraphics[width=\linewidth]{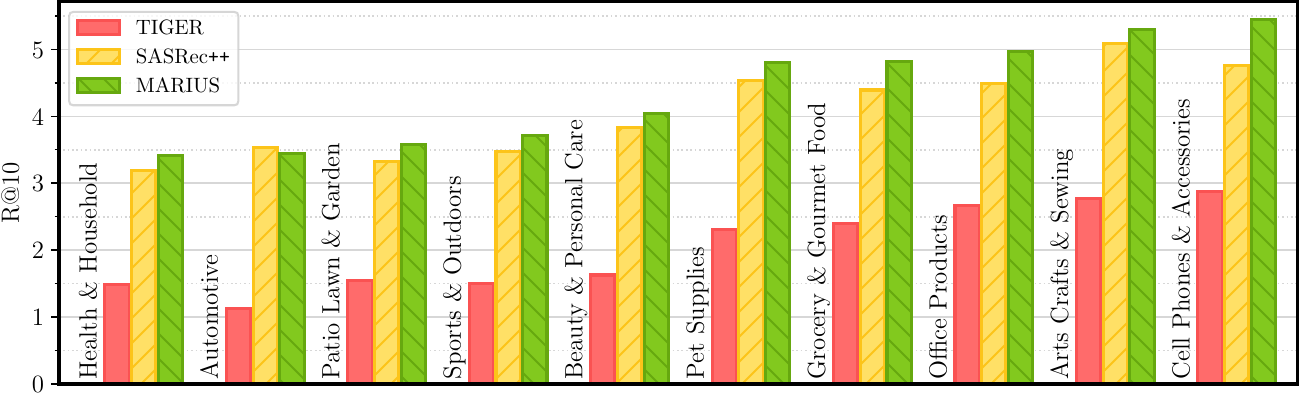}
                \caption{R@10 for large Amazon 2023 datasets. The number of users ranges from 197k (Arts, Crafts \& Sewing) to 796k (Health \& Household), while the number of items ranges from 77k (Office Products) to 267k (Automotive). While TIGER performs poorly at this scale, MARIUS and COSETTE demonstrate their effectiveness by outperforming our strong ID-based \SRpp{}.}
                \label{fig:AR23}
                \Description[Comparison of R@10 for TIGER, \SRpp{} and MARIUS]{For all datasets, MARIUS performs around twice as well as TIGER, and similarly or better than \SRpp{}.}
            \end{figure*}
        
            \begin{figure*}[t]
                \centering
                \begin{subfigure}[t]{0.32\linewidth}
                    \includegraphics[width=\linewidth]{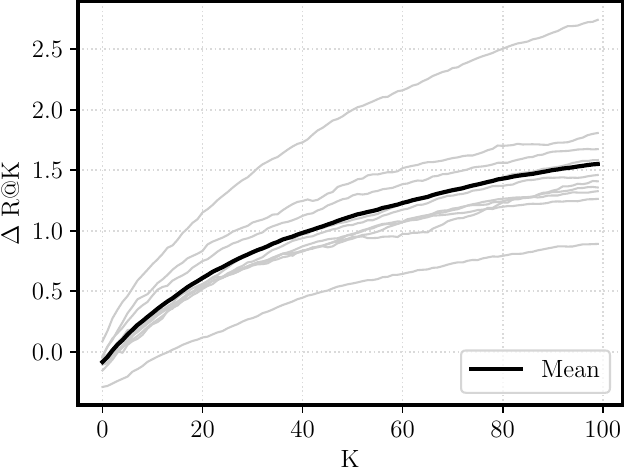}
                    \caption{Difference in R@K for varying K.}
                    \label{fig:recall}
                    \Description[Delta of Recall between \SRpp{} and MARIUS]{The mean delta starts around 0 for K=0, and reaches +1.5\% for K=100.}
                \end{subfigure}
                \hfill{}
                \begin{subfigure}[t]{0.32\linewidth}
                    \includegraphics[width=\linewidth]{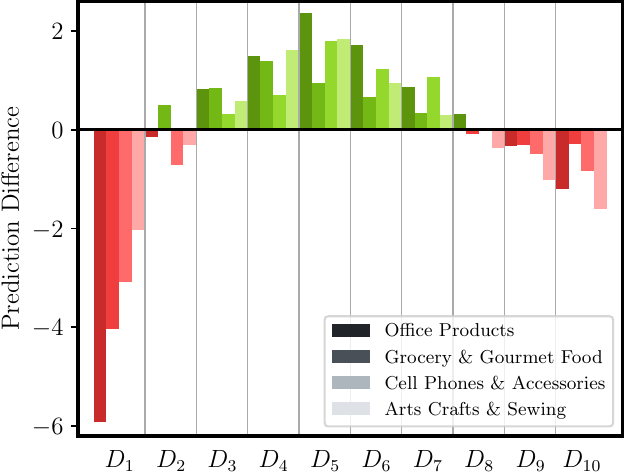}
                    \caption{Prediction difference in \% for each decile.}
                    \label{fig:generated_by_pop}
                    \Description[Delta of number of generation for each decile between \SRpp{} and MARIUS]{For 4 datasets, MARIUS predicts less items from the first decile, and more items from the deciles 3-7.}
                \end{subfigure}
                \hfill{}
                \begin{subfigure}[t]{0.32\linewidth}
                    \includegraphics[width=\linewidth]{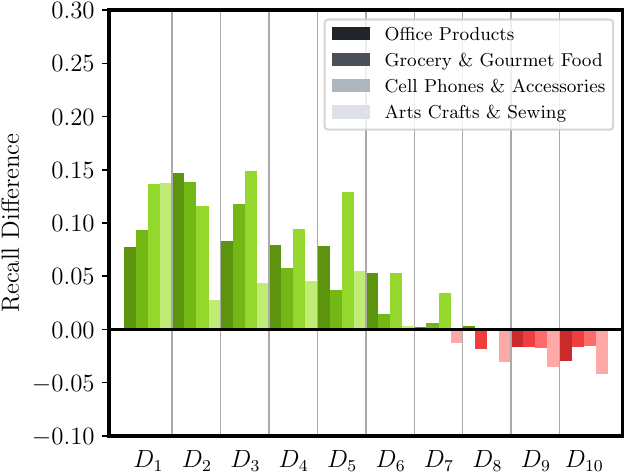}
                    \caption{Recall difference in \% for each decile.}
                    \label{fig:recall_by_pop}
                    \Description[Delta of R@10 for each decile between \SRpp{} and MARIUS]{For 4 datasets, MARIUS improves the recall over deciles 1-6.}
                \end{subfigure}
                \caption{\textbf{Performance analysis between MARIUS and \SRpp{}}. \textit{Left}: Difference in recall at K for vaying values of K, with MARIUS taking the lead for larger values of K. \textit{Middle}: Difference in prediction occurrences depending on item frequency. MARIUS tends to predict very popular items (1st decile) less and mid-popular items (decile 3 to 7) more. \textit{Right}: Difference in recall depending on item frequency. MARIUS makes better predictions on the first 70\% most frequent items.}
            \end{figure*}

            We first evaluate our method on standard sequential recommendation benchmarks from Amazon Reviews 2014, including \textit{Beauty}, \textit{Sports \& Outdoors}, and \textit{Toys \& Games}. Despite their limited scale, these datasets remain widely used in the literature and provide a consistent point of comparison with prior work. Results are reported in Table~\ref{tab:AR14}.

            To ensure comparability, all generative models are trained using Sentence-T5-XL embeddings. While stronger performance can be achieved with more powerful text encoders such as NV-Embed-v2, we observe that its high embedding dimensionality ($4096$ dimensions) leads to quantization issues on such small datasets (only 10–18k items). For this reason, we reserve its use for the larger-scale experiments in the next section.
            
            Overall, MARIUS trained with a basic RQ-VAE already surpasses previous T5-based generative recommenders. Replacing the quantizer with COSETTE leads to further gains, bringing performance comparable with our strong \SRpp{} baseline. In particular, using COSETTE IDs yields top R@10 for both \textit{Beauty} and \textit{Sports}, while remaining within standard deviation on R@5. Although we do not outperform \SRpp{} across all metrics, these results show that our approach closes much of the performance gap, demonstrating that generative recommenders can be made competitive even on small-scale benchmarks.
            
        \subsubsection{Scaling to Large Benchmarks}
            
            Model sizes are selected via grid search, guided by preliminary sizing experiments detailed in Appendix~\ref{app:model_sizing}. For SASRec, we search over $d \in \{128, 256\}$ and $L \in \{2, 4, 6\}$. For TIGER, we use an RQ-VAE with four quantization levels and apply distance-based deduplication to mitigate the high rate of tuple collisions observed with the default setting. We search over $d = 128$, $T_E \in \{2, 4, 6\}$, and $T_D \in \{6, 8, 10\}$. For MARIUS, we apply COSETTE for quantization and search over $d \in \{512, 768\}$, $T_E \in \{4, 6, 8\}$, and $T_D \in \{6, 8, 10\}$. All models are trained for 80k steps, and we report test performance for the best checkpoint selected by validation metrics.

            Results are summarized in Figure~\ref{fig:AR23}, with full tables provided in Appendix~\ref{tab:ar23_results}. While TIGER struggles at this scale and offers unsatisfactory performances, MARIUS combined with COSETTE consistently outperforms \SRpp{} across most datasets. We attribute these gains to the use of semantic information, which supports better generalization, as also observed by \cite{singh24better}.
            
        \subsubsection{Performance analysis}

            We conduct further analyses to understand which recommendation problems are most improved by MARIUS over \SRpp{}. 

            First, Figure~\ref{fig:recall} shows the difference of R@K between MARIUS and \SRpp{} for different values of K. For K $\leq 3$, MARIUS is very slightly under-performing. However, as K grows, MARIUS provides an increase recall gap. This behavior probably means that MARIUS is less focusing on the most probable item and is exploring alternative items. This is assumed to be a key feature of generative recommenders that are able to avoid the pure nearest-neighbor behavior of discriminative recommenders.

            We next analyze whether MARIUS is indeed exploring more alternative recommendations.
            Figure~\ref{fig:generated_by_pop} shows the difference in number of predictions between MARIUS and \SRpp{} depending on the occurrence frequency of items grouped by decile. We see that MARIUS predicts less the most frequent item occurrences (first decile) and much more the item occurrences from third to seventh decile. We attribute this to MARIUS being able to overcome the frequency bias in the data and learns that the most popular items are not necessarily always the ones to predict. This also comes at the cost of predicting less the rare items.

            To assess if this shift from most popular to mid-popular item is indeed beneficial, Figure~\ref{fig:recall_by_pop} shows the difference of recall between MARIUS and \SRpp{}. As we can see, the recall increases from the first decile to the seventh decile. This means that MARIUS is mostly able to exchange wrong highly popular item predictions for correct mid-popular items predictions. The performances on the most rare items decrease slightly, which means that rare event prediction is still very difficult.

        \subsubsection{Efficiency} \label{sec:efficiency}
            First, we report the asymptotic training and inference complexities of TIGER and MARIUS in Table~\ref{tab:complexities}, where $N$ is the sequence length, $L$ the semantic ID length, $T_E$ and $T_D$ are the transformer encoder and decoder costs, and $B$ is the number of beams to generate the items.
    
            Both models have comparable training complexities. Although TIGER includes an additional $L^2$ term in the encoder due to the semantic token input, this cost is small in practice since $L \ll N$. The main difference lies in the supervision strategy: MARIUS, being fully causal, allows the decoder to process and supervise all positions in parallel, at the same cost as TIGER supervising a single output because of the non-causal nature of the T5 encoder. In practice, all sequences are padded to the same length and we omit decoding over padding positions in MARIUS, which reduces the effective computational cost.
    
            At inference time, MARIUS is significantly more efficient. Unlike TIGER, which must attend to the full sequence for each generated item due to its cross-attention and encoder-decoder setup, MARIUS operates pointwise over the temporal context vectors. This makes its decoding cost independent of $N$. Furthermore, the bidirectional nature of TIGER’s T5 encoder prevents the use of KV-caching, further increasing its inference cost in practice.

            We then perform empirical measurements for both architectures, on a single H200, for models of similar size ($d=256$, 4 encoder and decoder layers). We use a batch size of 256 synthetic timelines, to remove the influence of the data loading, and average the results over 100 batches. 
            
            For training, we time a full optimization step and count the number of items over which the loss is computed. 
            For generation, we time a beam search of $B=10$ for each timeline, using a batched beam search implementation that processes all the beams in parallel.
            At a timeline length of 50, as used in this study, MARIUS trains on $33\times$ more items per second, and generates items $3\times$ faster.
            
            \begin{table}[t]
                \centering
                \caption{Training and Inference complexities. Notations: $N$ sequence length, $L$ semantic ID length, $T_E$ transformer encoder/temporal layers, $T_D$ transformer decoder/depth layers, $B$ number of generated items.}
                \resizebox{\linewidth}{!}{
                    \begin{tabular}{c|cc}
                        \toprule
                         Model  & \bld{Training} & \bld{Inference}  \\
                         \midrule
                         TIGER  & $(NL)^2T_E + (NL^2+L^2)T_D$ & $(NL)^2T_E + B(NL^2+L^2)T_D$ \\
                         MARIUS & $N^2T_E+ NL^2T_D$ & $N^2T_E + BL^2T_D$ \\
                         \bottomrule
                    \end{tabular}
                }
                \label{tab:complexities}
            \end{table}

            \begin{figure}[t]
                \centering
                \includegraphics[width=\linewidth]{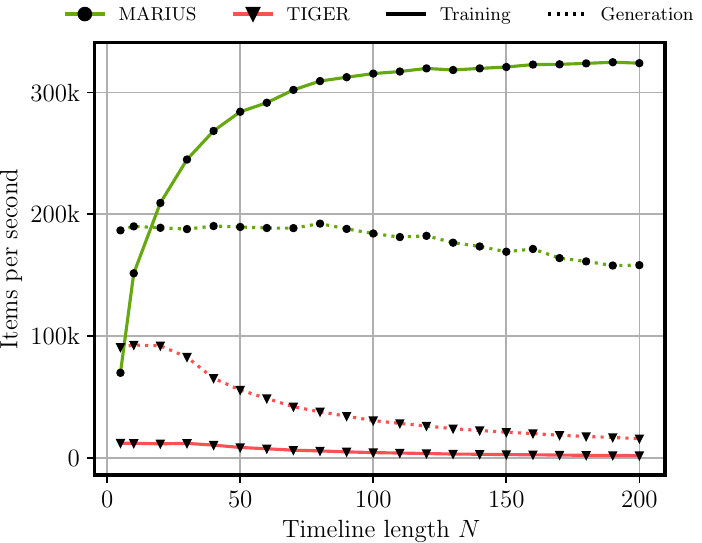}
                \caption{Speed comparison of MARIUS and TIGER. For training, we report the number of items used to compute the loss. For generation, we generate 10 items per timeline.}
                \label{fig:samples_per_second}
                \Description[Graph comparing MARIUS and TIGER's processing speeds]{The image is a line graph comparing the speed of two systems, MARIUS and TIGER, in terms of items processed per second over different timeline lengths (N). The x-axis represents the timeline length (N) ranging from 0 to 200, while the y-axis represents the number of items processed per second ranging from 0 to 300k.}
            \end{figure}

        Finally, regarding sampling efficiency, we find that top-10 predictions contain less hallucinations with our method compared to TIGER (see Table~\ref{tab:hallucinations} in the Appendix).
    \section{Conclusion}

We introduced COSETTE and MARIUS, two complementary components for improving generative recommendation systems. COSETTE improves the quality and uniqueness of semantic item representations by incorporating the downstream recommendation objective directly into the quantization process, resulting in more structured, task-aware semantic IDs than reconstruction-based alternatives. Building on these representations, MARIUS improves both the performance and computational efficiency of generation, particularly at scale. Together, COSETTE and MARIUS significantly reduce the performance gap between generative and discriminative recommenders, even surpassing them significantly on large scale datasets, making generative recommendation a more practical and competitive solution.
    
    {
        \bibliographystyle{acm}
        \bibliography{bib}
    }

    \clearpage

    \appendix

\section*{Appendix}

\section{COSETTE}
    \subsection{Parameters}
        \label{app:cosette_params}
        
\begin{figure}[b]
            \centering
            \includegraphics[width=.8\linewidth]{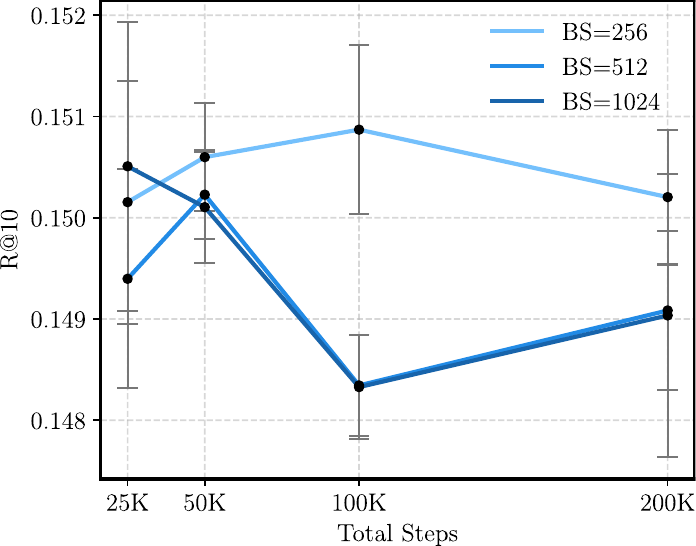}
            \caption{Impact of batch size and number of training steps on COSETTE performance (Video Games dataset). Smaller batch sizes lead to better generalization.}
            \label{fig:viga_BSxSteps}
            \Description[Graph showing COSETTE performance with varying batch sizes and training steps]{The image shows a graph depicting the impact of batch size and number of training steps on COSETTE performance using the Video Games dataset. The x-axis represents total steps (25K, 50K, 100K, 200K), and the y-axis represents R@10 values ranging from 0.148 to 0.152.}
        \end{figure}

\begin{figure}[t]
            \centering
            \includegraphics[width=.945\linewidth]{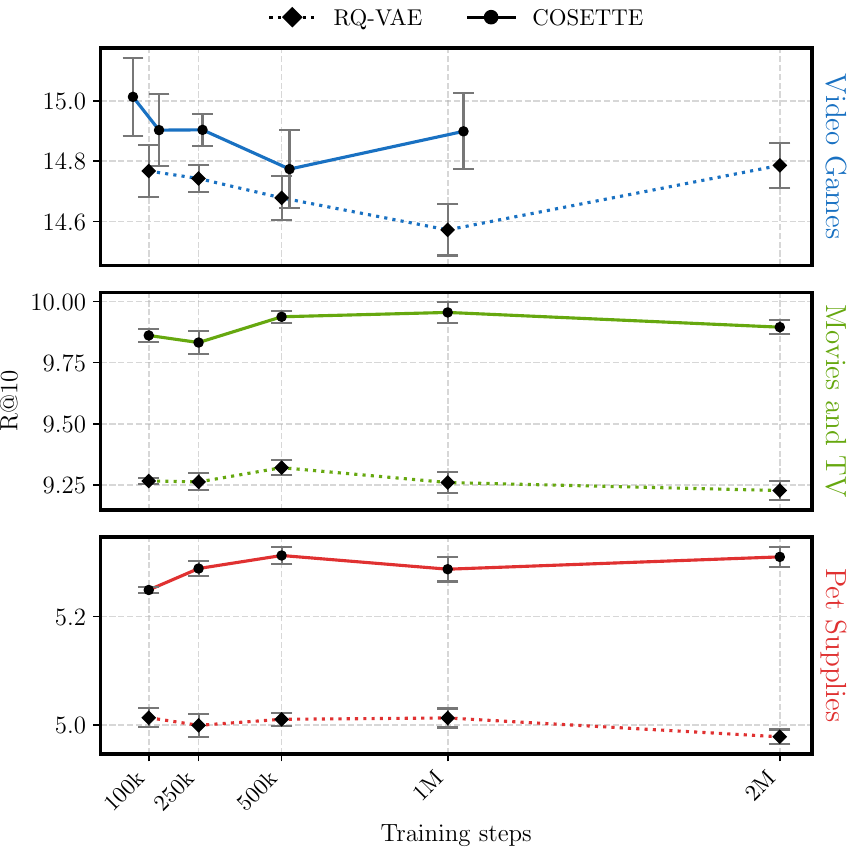}
            \caption{Validation R@10 of MARIUS models trained on embeddings from RQ-VAE and COSETTE, as a function of the quantizer’s training steps. We use a batch size of 1024. COSETTE benefits from extended training on large datasets.}
            \label{fig:cosette_steps}
            \Description[Three graphs comparing RQ-VAE and COSETTE performance across datasets]{The image shows three line graphs comparing the performance of RQ-VAE and COSETTE models on different datasets (Video Games, Movies and TV, Pet Supplies) as a function of training steps. The y-axis represents R@10 values, while the x-axis represents training steps (100k, 250k, 500k, 1M, 2M). Each graph shows how validation R@10 evolves for both models.}
        \end{figure}

        We analyze the sensitivity of COSETTE to several key hyperparameters across datasets of varying sizes. Our goal is to understand which configurations help stabilize training and improve performance, especially in low-resource regimes. All results are reported on validation sets, and summarized in Table~\ref{tab:cosette_params}, with additional trends illustrated in Figures~\ref{fig:cosette_steps} and~\ref{fig:viga_BSxSteps}.
    
        \paragraph{Loss weight.} 
        The contrastive loss in COSETTE is based on a binary cross-entropy formulation and typically reaches much larger values than the MSE-based reconstruction and quantization losses, which often converge to values on the order of $10^{-3}$. As a result, using a large $\lambda$ can cause the collaborative term to dominate training, leading to degraded quantization quality and inflated reconstruction errors. We find that setting $\lambda = 10^{-3}$ provides a good balance between the loss components. See Table~\ref{tab:cosette_params}a.
     
        \paragraph{Dropout.} 
        We apply dropout to both the encoder and decoder layers of the quantizer. This regularization proves especially beneficial on the smaller \textit{Video Games} dataset, which contains only 10k items, where overfitting is more likely. On larger datasets, the effect is smaller but still positive. These findings support the use of dropout as a regularization strategy in low-data regimes. See Table~\ref{tab:cosette_params}b.
        
        \paragraph{Temperature \& Bias.} 
        Although the temperature and bias in our contrastive loss are learned during training, we observe that their initialization has an impact on convergence and early performance. Empirically, we find that most datasets converge to temperature values around $t' = 2$ and bias values around $b = -8$. Initializing close to these values yields better results than neutral initializations (e.g., $t=1$, $b=0$), especially on large datasets. See Table~\ref{tab:cosette_params}c.
    
        \begin{table}[t]
            \centering
            \caption{Sensitivity of COSETTE to key hyperparameters across datasets. We report the mean R@10 over 3 runs for each configuration.}
            {\small
                \subfloat[Loss weight $\lambda$]{
    \begin{tabular}{c|ccc}
    \toprule
        $\lambda$  & Video Games & Movies \& TV & Pet Supplies \\
    \midrule
        $10^{-1}$ & 14.86       & 9.70       & 5.12 \\
        $10^{-2}$ & \bld{14.89} & 9.61       & 5.15 \\
        $10^{-3}$ & 14.77       & \bld{9.71} & \bld{5.17} \\
    \bottomrule
    \end{tabular}
}

\subfloat[Dropout]{
    \begin{tabular}{c|ccc}
    \toprule
        Dropout  & Video Games & Movies \& TV & Pet Supplies \\
    \midrule
        0   & 14.67       & 9.64       & 5.20 \\
        0.1 & \bld{14.89} & \bld{9.68} & \bld{5.21} \\
    \bottomrule
    \end{tabular}
}

\subfloat[Temperature \& Bias initialization]{
    \begin{tabular}{cc|ccc}
    \toprule
       $t'$ & $b$ & Video Games & Movies \& TV & Pet Supplies \\
    \midrule
        0 & 0  & 14.76       & 9.45       & 5.10 \\
        2 & 0  & \bld{14.87} & 9.47       & 5.13 \\
        0 & -8 & 14.52       & 9.65       & 5.18 \\
        2 & -8 & 14.72       & \bld{9.67} & \bld{5.20} \\
    \bottomrule
    \end{tabular}
}
            }
            \label{tab:cosette_params}
        \end{table}
        
        \paragraph{Batch size \& Steps.} 
        We investigate the effect of training length and batch size on the final quality of the learned semantic IDs. On larger datasets, model performance continues to improve steadily up to 500k steps (Figure~\ref{fig:cosette_steps}). However, on smaller datasets, prolonged training leads to overfitting. In these cases, using smaller batch sizes acts as a form of implicit regularization by increasing diversity in the target co-occurrences matrices. As shown in Figure~\ref{fig:viga_BSxSteps}, shorter training with smaller batches improves generalization in such regimes.

    \subsection{Collisions }
        
        Generative recommendation models rely on uniquely assigned Semantic IDs to accurately retrieve items during inference. However, standard RQ-VAE models do not explicitly enforce this uniqueness constraint, as their training objective focuses solely on reconstructing the input space. This often results in distinct items being mapped to identical codes, which undermines retrieval accuracy.
        To mitigate this issue, prior work typically applies post hoc disambiguation techniques, such as appending non-semantic deduplication tokens or performing distance-based reallocation. While effective to some extent, these methods are heuristic, non-differentiable, and ideally should be avoided or minimized.
        
\begin{figure}[t]
            \centering
            \includegraphics[width=.9\linewidth]{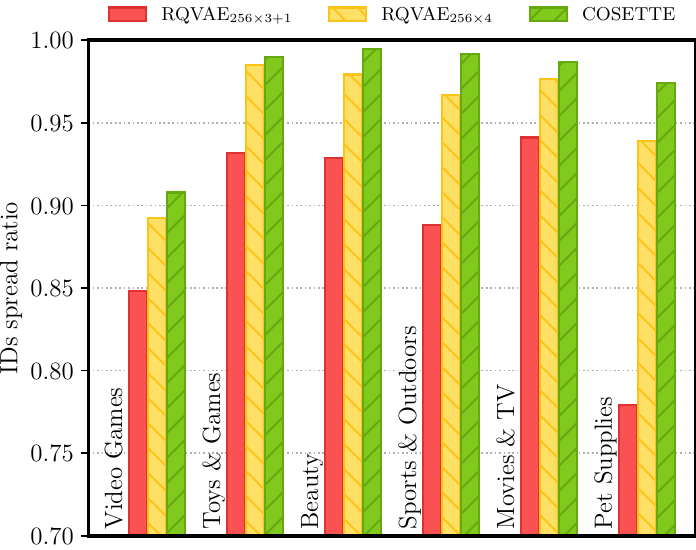}
            \caption{Ratio between the number of different IDs and the total number of items, before deduplicating procedures. The contrastive nature of COSETTE makes it produce more varied IDs.}
            \label{fig:collisions}
            \Description[Bar graph comparing ID diversity across methods and categories]{The image shows a bar graph comparing the ratio between the number of different IDs and the total number of items across various categories before deduplication procedures. The categories include Video Games, Toys & Games, Beauty, Sports & Outdoors, Movies & TV, and Pet Supplies. The graph compares three methods: RQVAE\textsubscript{256×3+1} (red), RQVAE\textsubscript{256×4} (yellow), and COSETTE (green).}
        \end{figure}

        Figure~\ref{fig:collisions} reports the ratio of unique semantic IDs produced after training relative to the total number of items in the dataset. As expected, increasing the number of quantization levels reduces the collision rate. COSETTE further reduces the collisions, owing to the collaborative contrastive loss which encourages the model to assign distinct codes to semantically similar items.

\section{MARIUS}
    \subsection{Architecture dimensions.}

        We investigate the impact of key architectural parameters of the MARIUS model on validation performance, using the Amazon 2014 datasets (\textit{Beauty}, \textit{Sports \& Outdoors}, and \textit{Toys \& Games}). Specifically, we ablate the number of layers, hidden dimensions, and dropout rates for both the Temporal and Depth Transformers. All experiments are conducted using a standard RQ-VAE quantizer.

        Starting from the default configuration described in Table~\ref{tab:abl_default_settings}, we vary each parameter independently to isolate its effect. The corresponding results are reported in Table~\ref{tab:ar14_ablation}.

        We observe largely consistent trends across the three datasets, with one notable exception: the optimal number of temporal layers appears to be slightly dataset-dependent. Based on these results, we select a configuration of $d=256$, 2 temporal layers, 2 depth layers, a dropout rate of 0.4 for the Temporal Transformer, and 0.1 for the Depth Transformer for subsequent experiments on the Amazon 2014 datasets.
                
        \begin{table}[t]
            \centering
            \caption{Default settings for our ablation study.}
            {\small
            \begin{tabular}{c|c|c}
                    \toprule
                    \multirow{3}{*}{Temporal} & Layers & 2 \\
                                              & Dimension & 256 \\
                                              & Dropout & 0.4 \\
                    \midrule
                    \multirow{3}{*}{Depth} & Layers & 2 \\
                                           & Dimension & 256 \\
                                           & Dropout & 0.1 \\
                    \midrule
                    & LR & $5e^{-4}$ \\
                    & WD & $1e^{-4}$ \\
                    & Steps & 70k \\
                    & Scheduling & Cosine \\
                    & Quantization & RQ-VAE 128 \\
                    & Model & Sentence-T5-XL \\
                    \bottomrule
                \end{tabular}
            }
            \label{tab:abl_default_settings}
        \end{table}
        \begin{table}[t]
            \centering
            \caption{Ablation study on the main parameters of MARIUS. We start from the default setting from Table~\ref{tab:abl_default_settings} and vary hyper-parameters one by one. We report the validation R@10.}
            {\small
                \subfloat[Temporal Layers]{
    \begin{tabular}{c|ccc}
    \toprule
          & Beauty & Sports & Toys \\
    \midrule
        1 & 10.90      & 6.60      & 10.14 \\
        2 & \bf{11.27} & 6.98      & 10.32 \\
        3 & 11.23      & \bf{7.06} & 10.29 \\
        4 & 11.09      & 6.82      & \bf{10.54} \\
        5 & 11.05      & 7.00      & 10.46 \\
    \bottomrule
    \end{tabular}
}
~
\subfloat[Depth Layers]{
    \begin{tabular}{c|ccc}
    \toprule
          & Beauty & Sports & Toys \\
    \midrule
        1 & 11.05      & 6.69      & 9.99 \\
        2 & \bf{11.18} & \bf{6.93} & \bf{10.46} \\
        3 & 11.08      & 6.81      & 10.00 \\
        4 & 10.61      & 6.77      & 9.92 \\
        5 & 10.55      & 6.76      & 9.54 \\
    \bottomrule
    \end{tabular}
}

\vspace{3mm}

\subfloat[Temporal Dimension]{
    \begin{tabular}{c|ccc}
    \toprule
            & Beauty & Sports & Toys \\
    \midrule
        64  &  9.82      & 5.86      & 8.84 \\
        128 & 10.84      & 6.59      & 9.70 \\
        256 & \bf{11.17} & \bf{7.04} & 10.40 \\
        384 & 10.93      & 6.86      & \bf{10.41} \\
        512 & 10.73      & 6.47      & 10.40 \\
    \bottomrule
    \end{tabular}
}
~
\subfloat[Depth Dimension]{
    \begin{tabular}{c|ccc}
    \toprule
            & Beauty & Sports & Toys \\
    \midrule
        64  &  8.77      & 5.05      & 7.24 \\
        128 & 10.23      & 6.05      & 9.29 \\
        256 & \bf{11.05} & \bf{6.96} & \bf{10.46} \\
        384 & 10.74      & 6.70      & 9.62 \\
        512 &  9.96      & 6.46      & 9.01 \\
    \bottomrule
    \end{tabular}
}

\vspace{3mm}

\subfloat[Temporal Dropout]{
    \begin{tabular}{c|ccc}
    \toprule
            & Beauty & Sports & Toys \\
    \midrule
        0.0 &  6.56      & 4.08      & 5.30 \\
        0.1 &  9.29      & 5.47      & 8.81 \\
        0.2 & 10.36      & 6.43      & 9.66 \\
        0.3 & 11.00      & 6.88      & 10.30 \\
        0.4 & \bf{11.03} & \bf{7.06} & \bf{10.40} \\
        0.5 & 10.94      & 6.74      & 10.08 \\
        0.6 & 10.21      & 6.36      & 9.59 \\
    \bottomrule
    \end{tabular}
}
~
\subfloat[Depth Dropout]{
    \begin{tabular}{c|ccc}
    \toprule
            & Beauty & Sports & Toys \\
    \midrule
        0.0 &  9.36      & 5.66      & 8.54 \\
        0.1 & \bf{11.04} & \bf{7.01} & \bf{10.41} \\
        0.2 & 10.89      & 6.47      & 9.82 \\
        0.3 &  9.70      & 5.79      & 8.82 \\
        0.4 &  8.99      & 5.26      & 7.97 \\
        0.5 &  8.26      & 4.77      & 6.74 \\
        0.6 &  7.62      & 4.37      & 5.71 \\
    \bottomrule
    \end{tabular}
}

            }
            \label{tab:ar14_ablation}
        \end{table}

    \begin{table*}[t]
        \centering
        \caption{Detailed results summarized in Figure~\ref{fig:AR23}. MARIUS achieves the highest R@10 on all but one dataset.}
        \resizebox{\linewidth}{!}{
            \begin{tabular}{l|cccc|cccc|cccc}
\toprule
 & \multicolumn{4}{c}{TIGER} & \multicolumn{4}{c}{SASRec\texttt{++}} & \multicolumn{4}{c}{MARIUS} \\
 & R@5 & NDCG@5 & R@10 & NDCG@10 & R@5 & NDCG@5 & R@10 & NDCG@10 & R@5 & NDCG@5 & R@10 & NDCG@10 \\
\midrule
Arts Crafts \& Sewing      & 1.75 & 1.14 & 2.77 & 1.47 & \textbf{3.51} & \textbf{2.42} & 5.09 & 2.93 & 3.49 & 2.37 & \textbf{5.3} & \textbf{2.95} \\
Automotive                 & 0.69 & 0.44 & 1.13 & 0.58 & \textbf{2.55} & \textbf{1.84} & \textbf{3.53} & \textbf{2.15} & 2.32 & 1.58 & 3.45 & 1.94 \\
Beauty \& Personal Care    & 0.98 & 0.64 & 1.63 & 0.84 & 2.68 & \textbf{1.88} & 3.84 & \textbf{2.25} & \textbf{2.71} & 1.81 & \textbf{4.04} & 2.24 \\
Cell Phones \& Accessories & 1.85 & 1.22 & 2.88 & 1.55 & 3.30 & 2.26 & 4.76 & 2.74 & \textbf{3.72} & \textbf{2.52} & \textbf{5.45} & \textbf{3.07} \\
Grocery \& Gourmet Food    & 1.52 & 1.00 & 2.39 & 1.28 & 3.02 & 2.07 & 4.39 & 2.51 & \textbf{3.22} & \textbf{2.15} & \textbf{4.83} & \textbf{2.67} \\
Health \& Household        & 0.96 & 0.64 & 1.49 & 0.81 & 2.20 & 1.53 & 3.19 & 1.85 & \textbf{2.27} & \textbf{1.54} & \textbf{3.42} & \textbf{1.91} \\
Office Products            & 1.81 & 1.25 & 2.67 & 1.53 & 3.25 & 2.33 & 4.49 & 2.73 & \textbf{3.49} & \textbf{2.43} & \textbf{4.98} & \textbf{2.91} \\
Patio Lawn \& Garden       & 0.90 & 0.58 & 1.54 & 0.78 & 2.37 & \textbf{1.68} & 3.32 & 1.99 & \textbf{2.43} & 1.65 & \textbf{3.58} & \textbf{2.02} \\
Pet Supplies               & 1.41 & 0.90 & 2.31 & 1.19 & 3.05 & 2.08 & 4.53 & 2.56 & \textbf{3.16} & \textbf{2.11} & \textbf{4.81} & \textbf{2.64} \\
Sports \& Outdoors         & 0.94 & 0.61 & 1.50 & 0.80 & 2.41 & \textbf{1.69} & 3.48 & 2.03 & \textbf{2.47} & 1.66 & \textbf{3.72} & \textbf{2.06} \\
\bottomrule
\end{tabular}
        }
        \label{tab:ar23_results}
    \end{table*}
    
    \subsection{Hallucinations}
    
        Since we do not impose constraints during beam search, the models may generate tuples that do not correspond to any existing item in the catalog. However, we observe that the frequency of such invalid generations is already low for TIGER, and is further reduced when using our method. This confirms that constrained decoding is not necessary in practice, as the models naturally learn to generate mostly valid item identifiers.

        \begin{table}[t]
            \centering
            \caption{Number of hallucinated items for MARIUS and TIGER for Top-10 predictions over the entire dataset. The hallucination rate for TIGER is very low, but MARIUS reduces it further.}
            \label{tab:hallucinations}
            \begin{tabular}{r|ccc}
   \toprule
        & \bld{TIGER} & \bld{MARIUS} & \light{Total} \\
    \midrule
    Video Games  &  87 & 7   & \light{243,030} \\
    Movies \& TV & 895 & 119 & \light{1,239,600} \\
    Pet Supplies & 261 & 14  & \light{5,948,000} \\
   \bottomrule
\end{tabular}
        \end{table}

\section{Model sizing}
    \label{app:model_sizing}

    \balance 
        
     To avoid extensive hyperparameter searches on each dataset, we train a range of models on datasets of varying scales: \textit{Beauty} (2014, 12k products), \textit{Office Products} (2023, 77k products), and \textit{Beauty \& Personal Care} (2023, 207k products). 

    \subsection{\SRpp{}}
       We study the behavior of \SRpp{} with respect to the model dimension and number of layers. All models use a $0.4$ dropout, and are trained with a maximum learning rate of $5 \cdot 10^{-4}$ using linear warmup and cosine annealing over 80k steps with Adam. We select the best checkpoint based on the validation split and report the mean HR@10 over three runs in Figure~\ref{fig:sasrec_size}.

        For the two larger datasets, we achieve better performance and stability by applying the InfoNCE loss with a temperature of $0.05$ against 30k randomly sampled negatives. For the smallest dataset, better results are obtained without normalization, though we observe strong overfitting and instability in the largest models. Overall, larger datasets benefit from slightly larger models.

    \subsection{MARIUS}

        We study the behavior of MARIUS with respect to model size across three dimensions: hidden size (shared across both scales), number of temporal layers, and number of depth layers. Semantic IDs are constructed from Sentence-T5-XL embeddings using an RQ-VAE with a latent dimension of $128$ and three codebooks containing $256$ centroids each, with an additional non-semantic deduplicatoin token. We fix the sequence length at $50$, the temporal dropout at $0.4$, the depth dropout at $0.1$, and the head dimension at $64$. Models are trained with AdamW for 80k steps using a batch size of $256$, a learning rate of $5 \cdot 10^{-4}$ and a weight decay of $10^{-4}$, with linear warmup and cosine annealing.
        
\begin{figure*}[p]
        \centering
        \includegraphics[width=.7\linewidth]{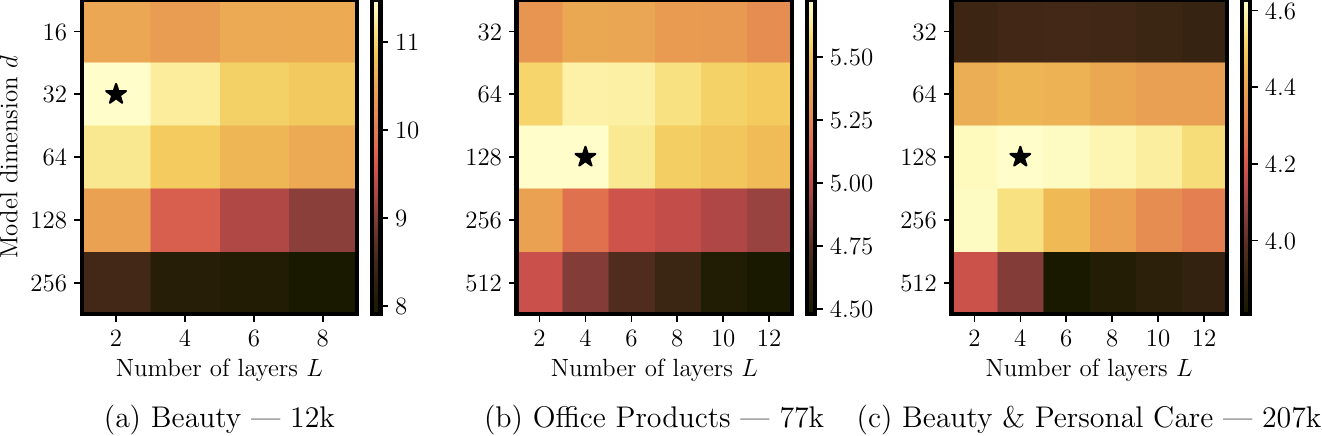}
        \caption{Mean validation R@10 across 3 runs of \SRpp{}, for different sizes of datasets. Note that the number of parameters in the embedding table grows linearly with the number of items. The star highlights the best configuration for each dataset.}
        \label{fig:sasrec_size}
        \Description[Heatmaps of R@10 results of \SRpp{}.]{Heatmaps of R@10 results of \SRpp{} across datasets (\textit{Beauty}, \textit{Office Products}, \textit{Beauty \& Personal Care}), dimensions and number of layers (2, 4, 6, 8). Optimality is reached for $d=32, L=2$ for \textit{Beauty}, $d=128, L=4$ for \textit{Office Products}, and $d=128, L=4$ for \textit{Beauty \& Personal Care}}            
    \end{figure*}

        We present the results in Figure~\ref{fig:marius_size} and analyze the optimal scaling strategy for MARIUS. The hidden dimension should increase with dataset size. The temporal transformer behaves similarly to SASRec, modeling sequences with a single token per item, and reaches optimal performance with four layers, which is comparable to Figure~\ref{fig:sasrec_size}. The depth transformer must scale with the number of items, as it stores the generative index responsible for tuple generation.

    \subsection{TIGER}

\begin{figure*}[p]
        \centering
        \begin{minipage}{.54\textwidth}
          \centering
          \includegraphics[width=\linewidth]{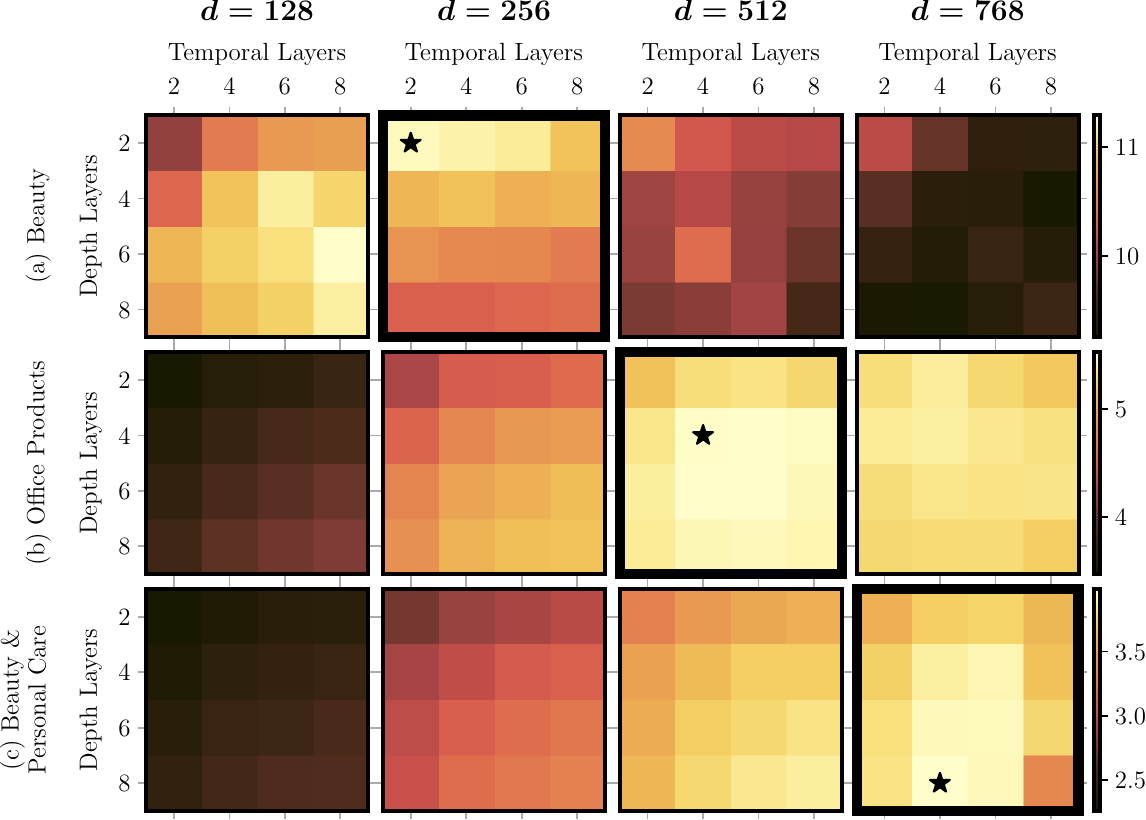}
          \captionof{figure}{Validation R@10 of MARIUS on RQ-VAE indices, for different datasets and model sizes. Cells with bold outlines report the mean of 3 runs. Stars highlight the selected configuration for each dataset.}
          \label{fig:marius_size}
          \Description[Heatmaps of R@10 results of MARIUS.]{Heatmaps of R@10 results of MARIUS across datasets (\textit{Beauty}, \textit{Office Products}, \textit{Beauty \& Personal Care}), dimensions (128, 256, 512, 768) and temporal and depth layers (2, 4, 6, 8). Optimality is reached for $d=256, L_T=2, L_D=2$ for \textit{Beauty}, $d=512, L_T=4, L_D=4$ for \textit{Office Products}, and $d=768, L_T=4, L_D=8$ for \textit{Beauty \& Personal Care}}
        \end{minipage}\hfill{}
        \begin{minipage}{.43\textwidth}
            \centering
            \includegraphics[width=\linewidth]{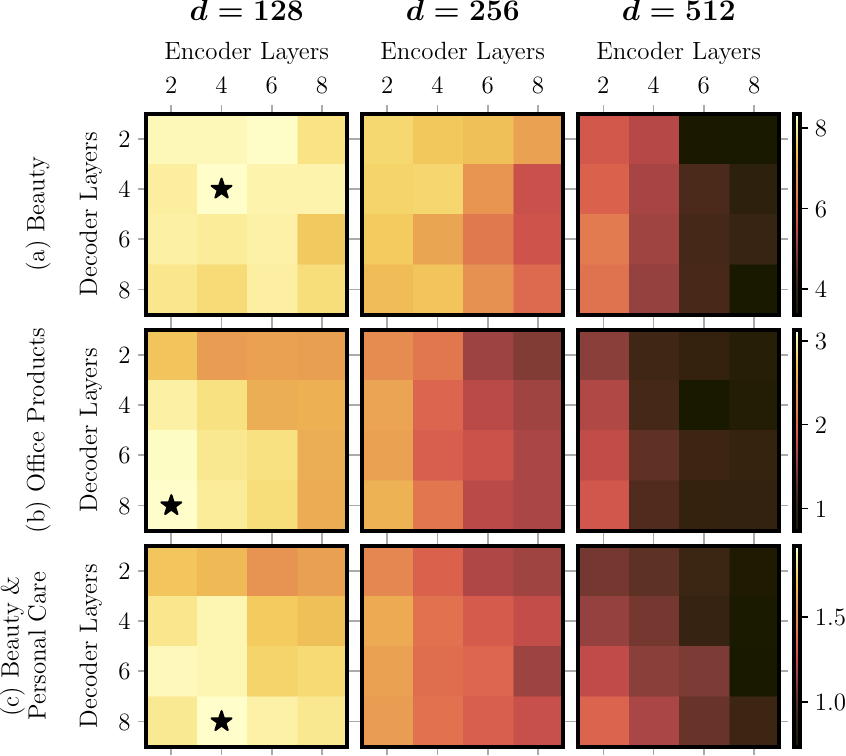}
            \captionof{figure}{Validation R@10 of TIGER applied on RQ-VAE indices, for different datasets and model sizes. Stars highlight the best configuration for each dataset.}
            \label{fig:tiger_size}
            \Description[Heatmaps of R@10 results of TIGER.]{Heatmaps of R@10 results of MARIUS across datasets (\textit{Beauty}, \textit{Office Products}, \textit{Beauty \& Personal Care}), dimensions (128, 256, 512) and encoder and decoder layers (2, 4, 6, 8). Optimality is reached for $d=128, L_E=4, L_D=4$ for \textit{Beauty}, $d=128, L_E=2, L_D=8$ for \textit{Office Products}, and $d=128, L_E=4, L_D=8$ for \textit{Beauty \& Personal Care}}
        \end{minipage}
    \end{figure*}

        We study the behavior of TIGER with respect to the number of encoder layers and decoder layers. The semantic IDs are the same as for MARIUS scaling in Figure~\ref{fig:marius_size}. Following the original paper, we fix the dropout rate at $0.1$ and the head dimension at $64$. Models are trained with AdamW for 80k steps using a batch size of $256$, and a learning rate of $5 \cdot 10^{-4}$, with linear warmup and cosine annealing.

        We present the results of scaling TIGER in Figure~\ref{fig:tiger_size}. Unlike MARIUS, TIGER performs best with a smaller hidden dimension (128), which aligns with its design: since item representations are not fused into a single token and stay available through cross-attention during the decoding, there is less need for high-dimensional embeddings. Similarly, the number of encoder layers should remain low, consistent with what is observed for both \SRpp{} and MARIUS. In contrast, the decoder must be scaled with the number of items, and TIGER appears to benefit from a greater number of decoder layers than MARIUS, likely due to the additional complexity introduced by the cross-attention mechanism.

    
\end{document}